\documentclass[twocolumn]{article}
\pdfoutput=1
\usepackage{graphicx,xcolor,dcolumn,bm,amsmath,amssymb,authblk,soul,abstract}
\usepackage{bbold}
\definecolor{dred}{rgb}{0.6,0,0}
\definecolor{dblue}{rgb}{0,0,0.6}
\definecolor{dgreen}{rgb}{0,0.6,0}
\usepackage[colorlinks=true,linkcolor=dblue,citecolor=dgreen]{hyperref}
\usepackage[top=1.5cm, bottom=2cm, left=2cm, right=1.2cm]{geometry}
\usepackage[bf,small]{titlesec}
\usepackage{aliascnt}
\usepackage{cleveref}

\title{Fast sampling for Bayesian inference in neural circuits}
\author[1]{Guillaume Hennequin\thanks{\small{\texttt{gjeh2@cam.ac.uk}}}}
\author[2]{Laurence Aitchison}
\author[1]{M\'at\'e Lengyel}
\affil[1]{Computational \& Biological Learning Lab, 
Department of Engineering, 
University of Cambridge, UK}
\affil[2]{Gatsby Computational Neuroscience Unit, 
University College London, UK}
\date{\bfseries This manuscript is a preliminary written version of
our Cosyne poster \cite{Hennequin14b}}

\renewcommand\b\mathbf
\renewcommand\d{{\rm d}}
\begin{document}

\twocolumn[
\maketitle
\begin{onecolabstract} 
Time is at a premium for recurrent network dynamics, and particularly so when
they are stochastic and correlated: the quality of inference from such dynamics
fundamentally depends on how fast the neural circuit generates new samples from
its stationary distribution. Indeed, behavioral decisions can occur on fast
time scales ($\sim 100$~ms), but it is unclear what neural circuit dynamics
afford sampling at such high rates.  We analyzed a stochastic form of
rate-based linear neuronal network dynamics with synaptic weight matrix
$\b{W}$, and the dependence on $\b{W}$ of the covariance of the stationary
distribution of joint firing rates. This covariance $\b\Sigma$ can be actively
used to represent posterior uncertainty via sampling under a linear-Gaussian
latent variable model. The key insight is that the mapping between $\b{W}$ and
$\b\Sigma$ is degenerate: there are infinitely many $\b{W}$'s that lead to
sampling from the same $\b\Sigma$ but differ greatly in the speed at which they
sample.  We were able to explicitly separate these extra degrees of freedom in
a parametric form and thus study their effects on sampling speed.  We show that
previous proposals for probabilistic sampling in neural circuits correspond to
using a symmetric $\b{W}$ which violates Dale's law and results in critically
slow sampling, even for moderate stationary correlations. In contrast,
optimizing network dynamics for speed consistently yielded asymmetric $\b{W}$'s
and dynamics characterized by fast transients, such that samples of network
activity became fully decorrelated over $\sim 10$~ms. Importantly, networks
with separate excitatory/inhibitory populations proved to be particularly
efficient samplers, and were in the balanced regime. Thus, plausible neural
circuit dynamics can perform fast sampling for efficient decoding and
inference.  
\end{onecolabstract}
\vspace*{1cm}
]
\saythanks

\section{Introduction}

\begin{figure}[!t] 
\centering
\includegraphics[width=0.9\columnwidth]{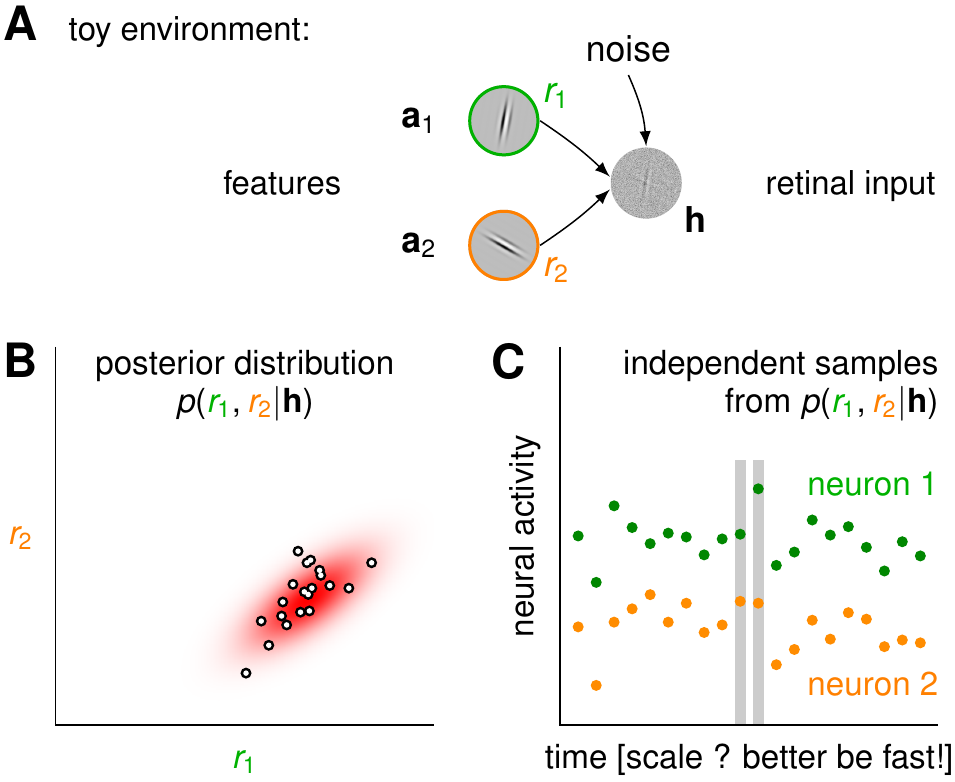}
\caption{\label{fig:sampling}{\bfseries Sampling-based representation
of (perceptual) uncertainty.} A toy visual environment (A) comprises two
features (oriented edges) $\mathbf{a}_1$ and $\mathbf{a}_2$ which are present
in the scene with intensities $r_1$ and $r_2$ respectively. The two features
combine linearly to form the ``retinal input'' $\b{h}$, to which noise is
added.  Perception is about inferring the intensities $r_1$ and $r_2$ at
which the features are present in the scene, given the retinal input
$\b{h}$. If the uncertainty about $r_1$ and $r_2$ matters, e.g. for making
optimal decisions when they too have uncertain consequences~\cite{Fiser10}, one
must represent the full distribution $p(r_1,r_2|\b{h})$ (B).  This can be
done by drawing (independent) samples from that distribution (B, white dots),
which the brain could encode in the joint activity of two neurons (C; each time
frame corresponds to one of the white dots in B). For sampling to be of any
practical use, the minimum time that separates the collection of two
independent samples (gray frames in C) must be short.}
\vspace*{0.3cm} 
\hrule
\end{figure}

Perception in humans is blazingly fast: when presented with an image for 20~ms,
we can tell in a split second whether or not it contained an animal, and our
brain holds the correct answer as early as 150~ms following stimulus
onset~\cite{Thorpe96}. Such celerity is surprising given the difficulty of the
task: sensory inputs being noisy and ambiguous (\Cref{fig:sampling}A), they
do not uniquely determine the state of the
environment, so perception is inherently a matter of probabilistic
inference~\cite{Knill04}.  Thus, the brain must represent and compute with
complex probability distributions over relevant environmental variables.
Most state-of-the-art machine learning techniques for solving similar inference
problems at large scale face a tradeoff between inference accuracy and
computing speed (e.g.~\cite{Mackay03}). The brain, on the contrary, seems to
enjoy both simultaneously.

Some probabilistic computations can be made easier through an appropriate
choice of representation for the probability distributions of interest.
Sampling-based representations
(\Cref{fig:sampling}B,~\cite{Fiser10,Berkes11}), for example, make computing
moments of the distribution or its marginals straightforward. However, the
speed issue (cf. above) becomes even more fundamental: no matter how close the
actual sampled distribution is to the ideal one, any sampling-based computation
becomes accurate only after enough samples have been collected, and one has no
choice but \emph{waiting} for those samples to be delivered by the circuit
dynamics.  For sampling to be of any practical use, the interval that separates
the generation of two independent samples must be short relative to the desired
behavioural timescale (\Cref{fig:sampling}C). Single neurons can integrate
their inputs on a timescale $\tau_{\rm m}\approx 10-50$~ms, whereas we must
often make decisions in less than a second: this leaves just enough time to use
(i.e. read out) a few tens of samples. Thus, it seems that the dynamics of
brain circuits cannot afford correlating neural activity on a timescale
longer than $\tau_{\rm m}$. How such temporal decorrelation can be achieved in
cortical circuits remains unclear.
 
In this note, we introduce a simple yet non-trivial generative model and seek
plausible neuronal network dynamics for \emph{fast} sampling from the
corresponding posterior distribution. While some standard machine learning
techniques do suggest ``neural network''-type solutions to sampling, not only
are the corresponding architectures implausible in fundamental ways (e.g.  they
violate Dale's law), but we show here that they lead to unacceptably slow
sampling in high dimensions. Although this problem is already well appreciated
in the machine learning community, the simplicity of our generative model
allows us to draw an analytical picture of it and to suggest solutions.  In
fact, we can use methods from robust control to discover the \emph{fastest}
neural-like sampler for our generative model, and study its structure. We find
that it corresponds to greatly non-symmetric synaptic interactions (in contrast
to most off-the-shelf samplers), and mathematically nonnormal circuit
dynamics~\cite{Trefethen05}, in striking agreement with our current
understanding of primary visual cortex (V1) dynamics~\cite{Murphy09}.


\section{Basic setup}

\begin{figure}[b]
\hrule
\vspace*{0.5cm}
\centering
\includegraphics[width=\columnwidth]{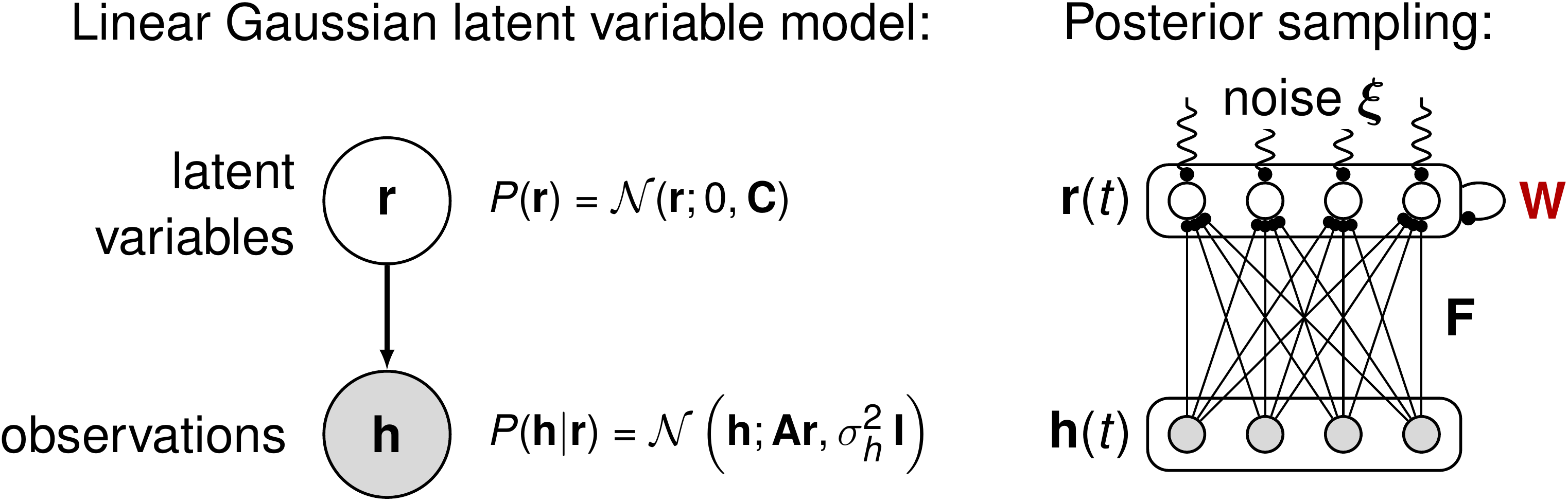}
\caption{\label{fig:schematics}{\bfseries Sampling under a linear Gaussian 
latent variable model using neuronal network dynamics.}
Left: schematics of the generative model. Right: schematics of the recognition
model. Sampling from $p(\b{r}|\b{h})$ is achieved through the linear, recurrent
processing of both the input $\b{h}$ and some private sources of noise $\boldsymbol\xi$
(see text). $\b{F}$ and $\b{W}$ denote feedforward and recurrent synaptic weight
matrices respectively.} 
\end{figure}

We focus on the linear Gaussian latent variable model, which is a
high-dimensional generalization of the example given in
\Cref{fig:sampling}A.  The model generates observations $\b{h} \in
\mathbb{R}^M$ as weighted sums of $N$ features
$\left(\b{a}_1,\ldots,\b{a}_N\right) \in \mathbb{R}^{M\times N}$ with
jointly Gaussian coefficients $(r_1,\ldots,r_N)$, plus independent additive
noise terms (\Cref{fig:schematics}, left).
More formally:
\begin{align}
p(\b{r}) \quad&=\quad \mathcal{N}(\b{r};0,\b{C})\\ 
p(\b{h}|\b{r}) \quad&=\quad \mathcal{N}\left(\b{h}; \b{A}\b{r}, \sigma_h^2 \b{I}\right)\\
\b{A} \quad&=\quad \left(\b{a}_1;\b{a}_2;\ldots;\b{a}_N\right)
\end{align}
The posterior distribution is multivariate Gaussian:
\begin{align}
p(\b{r}|\b{h}) \quad&=\quad \mathcal{N}\left(\b{r};\boldsymbol\mu(\b{h}),\b\Sigma\right)\\
\b\Sigma \quad&=\quad \left( \b{C}^{-1} + \frac{\b{A}^\top \b{A}}{\sigma_h^2}\right)^{-1}\\
\boldsymbol\mu(\b{h}) \quad&=\quad \b\Sigma \b{A}^\top \b{h} / \sigma_h^2.
\end{align}
We are interested in neural circuit dynamics for sampling from $p(\b{r}|\b{h})$,
whereby the data (observation) $\b{h}$ is given as a constant feedforward input
to another layer of recurrently connected units, which encode the latent variables
and also receive inputs from external, private sources of noise
$\boldsymbol\xi$ (\Cref{fig:schematics}, right).  The activity fluctuations
$\b{r}(t)$ in the recurrent layer must have a stationary distribution that
matches the posterior, for any $\b{h}$.

More precisely, we consider linear recurrent stochastic dynamics of the form:
\begin{equation}\label{eq:ouprocess}
\d\b{r} = \frac{\d t}{\tau_{\rm m}} (-\b{r}(t)+\b{W}\b{r}(t)+\b{F}\b{h}) 
  + \sigma_\xi \sqrt{\frac2{\tau_{\rm m}}}\: \d\boldsymbol{\xi}(t)
\end{equation}
Here $\tau_{\rm m}$ is the single-unit ``membrane'' time constant, and
$\d\boldsymbol{\xi}$ is a Wiener process of unit variance, which is scaled by a
scalar noise intensity $\sigma_\xi$. The activity $r_i(t)$ could represent
either the membrane potential of neuron $i$, or the deviation from baseline of
its momentary firing rate. The matrices $\b{F}$ and $\b{W}$ contain the
feedforward and recurrent connection weights, respectively.

The stationary distribution of $\b{r}$ is indeed Gaussian with a mean 
\begin{equation}
\boldsymbol\mu^{\b{r}}(\b{h}) = (\b{I}-\b{W})^{-1}\b{F}\b{h}
\end{equation}
and a covariance matrix $\b\Sigma^{\b{r}} \equiv \left\langle
(\b{r}(t)-\boldsymbol\mu^{\b{r}}) (\b{r}(t)-\boldsymbol\mu^{\b{r}})^\top
\right\rangle_t$ that depends only on $\b{W}$ and $\sigma_\xi$, but not on
$\b{h}$, according to the following Lyapunov equation \cite{Gardiner85}:
\begin{equation}\label{eq:lyap}
(\b{W}-\b{I})\b\Sigma^{\b{r}} + \b\Sigma^{\b{r}} (\b{W}-\b{I})^\top = - 2\sigma_\xi^2 \b{I} 
\end{equation}
where $\b{I}$ denotes the identity matrix. Note that in the absence of a
recurrent connectivity ($\b{W}=0$), the variance of every $r_i(t)$ would be
exactly $\sigma_\xi^2$.

In order for the dynamics of \Cref{eq:ouprocess} to sample from the right
posterior, we must choose $\b{F}$, $\b{W}$ and $\sigma_\xi$ such that
$\boldsymbol\mu^{\b{r}}(\b{h})~=~\boldsymbol\mu(\b{h})$ and $\b\Sigma^{\b{r}} =
\b\Sigma$. A possible combination is
\begin{align}
\b{F} \quad&=\quad \left(\sigma_\xi/\sigma_h\right)^2\: \b{A}^\top
\label{eq:langevinf}\\
\b{W} \quad&=\quad \b{W}_{\rm L} \quad\equiv\quad \b{I} - 
\sigma_\xi^2 \: \b\Sigma^{-1} \label{eq:langevinw}\\
\sigma_\xi \quad&{\rm arbitrary,} >0
\end{align}

In the study that follows, we will be interested in the likelihood matrix
$\b{A}$ only insofar as it affects the posterior covariance matrix $\b\Sigma$.
We can in fact ignore $\b{A}$ altogether, and focus on the case where
$\b{h}=0$, so that the posterior collapses to the prior with a covariance
matrix $\b\Sigma=\b{C}$ whose structure turns out to be the only thing that
affects the speed of sampling.

\section{\label{sec:langevin}Langevin sampling is very slow}

Langevin sampling \cite{Mackay03,Neal11,Welling11} is a common sampling
technique, in which a stochastic dynamical system performs a ``noisy gradient
ascent of the log posterior'':
\begin{equation}\label{eq:langevin}
\d\b{r} = \frac\partial{\partial\b{r}} \log\: p(\b{r}|\b{h}) \: \d t
  + \d \boldsymbol\xi 
\end{equation}
(where $\d \boldsymbol\xi$ is a unitary Wiener process). 
When $\b{r}|\b{h}$ is Gaussian, \Cref{eq:langevin} reduces to
\Cref{eq:ouprocess} for $\sigma_\xi=1$ and the choice of $\b{F}$ and
$\b{W}$ given in \Cref{eq:langevinf,eq:langevinw}
-- hence the notation $\b{W}_{\rm L}$. Note that $\b{W}^{\rm L} = \b{W}_{\rm
L}^\top$, i.e. it is a symmetric weight matrix.
         
\newcommand\wl{\b{W}_{\rm L}}
\newcommand\lmw{\lambda_{\rm max}^{\b{W}-\b{I}}}
\newcommand\lmwl{\lambda_{\rm max}^{\wl-\b{I}}}
\newcommand\lms{\lambda_{\rm max}^{\b\Sigma}}
\newcommand\lmsinv{\lambda_{\rm min}^{\b\Sigma^{-1}}}
 
As we show now, this choice of weight matrix leads to very slow mixing (i.e.
very long correlation times for $\b{r}(t)$) in any high-dimensional sampling
space ($N\gg 1$).  In a linear network, the average autocorrelation length is
dominated by the decay time constant $\tau_{\rm max}$ of the slowest
eigenmode, i.e.  the eigenvector of $(\b{W}-\b{I})$ associated with the
eigenvalue $\lmw$ which, of all the eigenvalues of $(\b{W}-\b{I})$, has the
largest real part (which must stil be negative, for stability reasons). The
contribution of that slowest eigenmode to the sample autocorrelation time is
roughly $\tau_{\rm max}=-\tau_{\rm m}/{\rm Re}\left(\lmw\right)$, so sampling becomes
very slow when ${\rm Re}\left(\lmw\right)$ approaches $0$ (from below). This
is, in fact, what happens with Langevin sampling as $N\to \infty$. To see this,
let us recall that $(\wl-\b{I})$ is real and symmetric, so its eigenvalues are
all real, and since $\wl-\b{I}=-\sigma_\xi^2 \b\Sigma^{-1}$ we can
write\footnote{For a non-singular matrix $\b{M}$, the eigenvalues of
$\b{M}^{-1}$ are the inverses of those of $\b{M}$; and since $\b\Sigma$ is a
positive definite covariance matrix, all its eigenvalues are positive, which
yields \Cref{eq:deriv1}.} 
\begin{equation}\label{eq:deriv1}
\lmwl \quad=\quad -\sigma_\xi^2\lmsinv \quad=\quad = -\frac{\sigma_\xi^2}\lms
\end{equation}
Now, again because of its symmetry, $\b\Sigma$ is a normal matrix, and so it is
is similar to (or equal to the unitary transformation of) a diagonal matrix
that contains its eigenvalues.  Since unitary transformations preserve the
Frobenius norm, we can write
\begin{equation}\label{eq:deriv2}
\sum_{i,j} \Sigma_{ij}^2 \quad = \quad  
\sum_i \left(\lambda_i^{\b\Sigma} \right)^2
\end{equation}
and since all the eigenvalues of $\b\Sigma$ are positive, 
\begin{equation}\label{eq:deriv3}
N \left(\lms\right)^2 \geq \sum_i \left(\lambda_i^{\b\Sigma}\right)^2
\end{equation}
Combining Eqs.~\ref{eq:deriv1}-\ref{eq:deriv3}, we arrive at a bound that
relates the maximum eigenvalue of $(\wl-\b{I})$ to a basic summary statistics
of the posterior covariance matrix $\b\Sigma$:
\begin{equation}\label{eq:genbound}
\lmwl \quad \geq \quad -\sigma_\xi^2\sqrt{\frac{N}{\sum_{ij} \Sigma^2_{ij}}}
\end{equation}
In the $N\to\infty$ limit, assuming that pairwise correlations do not vanish,
the denominator is expected to be $\mathcal{O}(N^2)$, meaning that $0>\lmwl
\geq -\mathcal{O}(1/\sqrt{N})$: the slowest eigenmode of $\wl$ becomes
critically slow for high-dimensional posteriors. To formalize this intuition,
let us assume that $\Sigma_{ii} \simeq \sigma_0^2$ (all posterior variances are
roughly equal) and that the distribution of pairwise posterior correlations has
zero mean and standard deviation $\sigma_r$. We can then rewrite
\Cref{eq:genbound} as
\begin{equation}\label{eq:finalbound}
\lmwl \quad \geq \quad \frac{-(\sigma_\xi/\sigma_0)^2}{\sqrt{1+N\sigma_r^2}}
\end{equation}

\begin{figure}[t]
\includegraphics[width=\columnwidth]{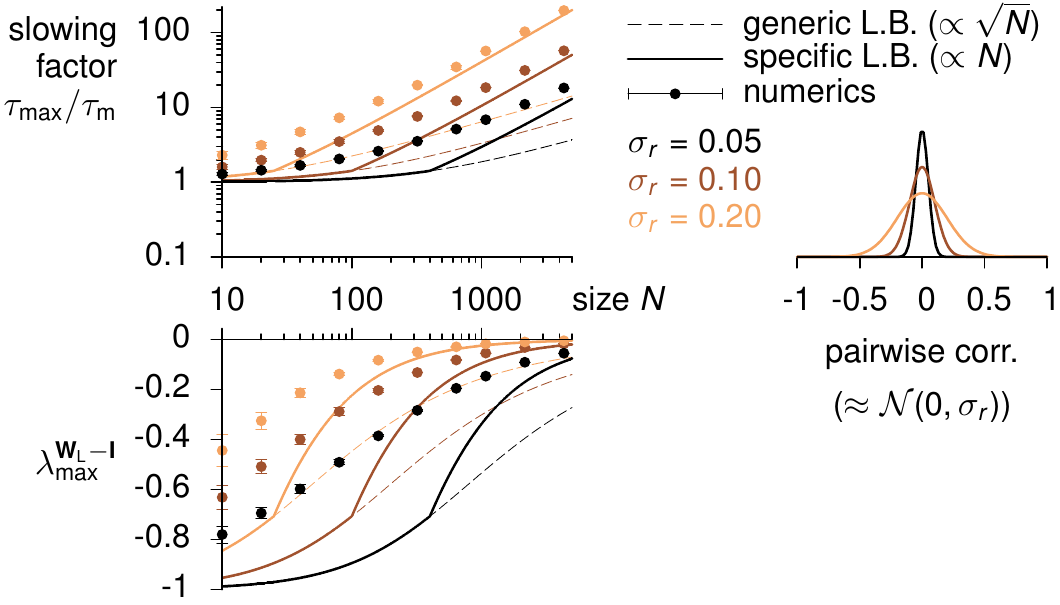}
\caption{\label{fig:scaling}{\bfseries Langevin sampling is slow in
high-dimension}. Random covariance matrices $\b\Sigma$ of size $N$ are drawn
from a Wishart distribution with $n$ degrees of freedom and scale matrix
$\sigma_0^2\b{I}/n$. This yields an average variance of $\sigma_0^2$, and
a distribution of pairwise correlations with zero mean and variance
$\sigma_r^2\approx 1/n$ (right).  Sampling from $\mathcal{N}(\cdot,\b\Sigma)$ using
a stochastic neural network (cf.  \Cref{fig:schematics}) with $\b{W}=\wl$
(Langevin, symmetric solution) becomes increasingly slow as $N$ grows, as
indicated by the relative decay time constant $\tau_{\rm max}/\tau_{\rm m}$ of
the slowest eigenmode of $(\wl-\b{I})$ (top), which is related to the inverse
of its largest eigenvalue (bottom). Dots indicate the numerical evaluation of
the corresponding quantities, for $100$ sample matrices for each $N$. Dashed
lines correspond to the generic bound in \Cref{eq:finalbound}. Solid lines are
refined bounds for the specific case $\b\Sigma \sim {\rm
Wishart}(\sigma_0^2\b{I}/n, n)$ with $n\approx 1/\sigma_r^2$
(\Cref{eq:wishartbound}). The two bounds merge for $N<n$.
Parameters were set to $\sigma_\xi=\sigma_0=1$.}
\vspace*{0.3cm}
\hrule 
\end{figure}

We see that Langevin sampling is bound to be slow in the limit of large state
spaces ($N \to \infty$) and when pairwise correlations do not vanish%
\footnote{Or vanish, but not as fast as $1/\sqrt{N}$.} in that limit
(\Cref{fig:scaling}, dashed lines).

We can refine this bound in the case when $\b\Sigma$ is drawn from a Wishart
distribution with $n$ degrees of freedom and scale matrix $(\sigma_0^2/n)\b{I}$
(\Cref{fig:scaling}). In this case, the expected value of a diagonal element
(variance) in $\b\Sigma$ is $\sigma_0^2$, and the distribution of
pairwise correlations is centered with variance $\sigma_r^2\approx 1/n$. If
$\sigma_r^2 \sim \mathcal{O}(1)$, $\b\Sigma$ becomes low-rank as $N$ grows,
and in fact, it has only $\approx \sigma_r^{-2}$ non-zero eigenvalues\footnote{In
principle, a singular $\b\Sigma$ would not be an appropriate posterior
covariance matrix -- and indeed, no linear stochastic network such as described
by \Cref{eq:ouprocess} would be able to sample from
$\mathcal{N}(\boldsymbol\mu,\b\Sigma)$ then. Introducing a small regularizer,
i.e. considering $\tilde{\b\Sigma} \equiv \b\Sigma + \varepsilon^2 \b{I}$,
solves the problem but does not alter the asymptotic properties of the bound
we derive here, so we omit this detail for the sake of clarity.}.
\Cref{eq:deriv3} can be adjusted to take this into account:
\begin{equation}\label{eq:wishartbound1}
\min(n,N) \left(\lms\right)^2
\geq \sum_i \left(\lambda_i^{\b\Sigma}\right)^2
\end{equation} 
As mentioned above, the r.h.s. of \Cref{eq:wishartbound1} is equal to the
squared Frobenius norm of $\b\Sigma$, which can be easily estimated for the
Wishart ensemble, at least for $N$ and $n$ not too small:
\begin{equation}
\| \b\Sigma \|_F^2  \quad\approx\quad N\sigma_0^2 \left( 1 +
 \frac{N}{n}\right)
\end{equation}
Using \Cref{eq:deriv1} and recalling that $n\approx \sigma_r^{-2}$, we
arrive at a Wishart-specific bound:
\begin{equation}\label{eq:wishartbound}
\lmwl \quad\geq\quad \frac{-(\sigma_\xi/\sigma_0)^2\sqrt{\min(\sigma_r^{-2},N)}}{\displaystyle
\sqrt{N(1+\sigma_r^2N)}}
\end{equation} 
Note that when $\sigma_r < 1/\sqrt{N}$, the bound becomes equivalent to the more
general case in \Cref{eq:finalbound}. For $\sigma_r\sim\mathcal{O}(1)$,
however, the slowing problem becomes worse for Wishart matrices, since
now $0>\lmwl\geq -\mathcal{O}(1/N)$ (\Cref{fig:scaling}, solid lines).

The ratio $(\sigma_0/\sigma_\xi$), which shows up in both versions of our bound
(\Cref{eq:finalbound,eq:wishartbound}), tells us how much the recurrent
interactions must amplify the external noise in order to produce from the right
stationary activity distribution (recall that $\sigma_\xi$ measures the
magnitude of the activity fluctuations due to the input noise alone, in the
absence of recurrent circuitry). The more amplification is required
($\sigma_\xi \ll \sigma_0$), the slower the dynamics of Langevin sampling is
bound to become.

In summary, Langevin sampling corresponds to symmetric interactions (which
violates Dale's law), and in the Gaussian case considered here, yields
unacceptably slow mixing in high-dimensional latent spaces. This should be true
whenever i) the magnitude of the posterior variances does not depend on $N$,
and ii) the spread of the distribution of posterior pairwise correlations also
does not depend on $N$.  The types of generative models under which the second
assumption holds are yet to be characterized; we leave this to future work.

\section{\label{sec:optimization}What is the fastest sampler?}

While Langevin dynamics (\Cref{eq:langevin}) give a general recipe for
sampling from any given posterior density, it unduly constrains the dynamics to
obey symmetric interactions -- at least in the Gaussian case.  To see why this
is a huge restriction, let us observe that \emph{any connectivity matrix of the
form}
\begin{equation}\label{eq:gen:solution}
\b{W}(\b{S}) = \b{I} + \left( - \sigma_\xi^2 \b{I} + \b{S} \right)
\b{\Sigma}^{-1}
\end{equation}
\emph{where $\b{S}$ is an arbitrary skew-symmetric matrix, solves
\Cref{eq:lyap}, and therefore induces the right stationary distribution
$\mathcal{N}(\cdot,\b\Sigma)$ under the linear stochastic dynamics of
\Cref{eq:ouprocess}}. Note that Langevin sampling (\Cref{eq:langevinw})
corresponds to $\b{S}=0$. But in general, there are $\mathcal{O}(N^2)$ degrees
of freedom in the skew-symmetric matrix $\b{S}$, which is a lot. Could these be
exploited to speed up mixing? In the following, we show that indeed a large
gain in sampling speed can be obtained through an appropriate choice of
$\b{S}$. By formulating the problem as one of robust control, we can use
optimization techniques to discover the \emph{fastest sampler}, in a sense that
we define below.


Let $\b{K}(\b{S},\tau) = \left\langle \delta\b{r}(t) \:
\delta\b{r}(t+\tau)^\top \right\rangle_t $ be the covariance matrix
between pairs of samples separated by a time interval $\tau$, in the stationary
regime (we use the notation $\delta\b{r}(t) \equiv \b{r}(t)-\boldsymbol\mu$).
Note that $\b{K}(\b{S},0)=\b\Sigma$ is the posterior correlation matrix, and
that for fixed $\sigma_\xi^2$ and $\tau_{\rm m}$, $\b{K}(\b{S},\tau)$ depends
only on the interval $\tau$ and on the matrix of recurrent weights $\b{W}$,
which itself depends only on our skew-symmetric free-parameter matrix $\b{S}$.
We define a ``total slowing cost''
\begin{equation}
\label{eq:slowingcost}
\psi_{\rm slow}(\b{S}) = \frac1{2\tau_{\rm m} N^2} 
  \int_0^\infty \left\| \b{K}(\b{S},\tau)
\right\|_{\rm F}^2\: \d\tau 
\end{equation}
which penalizes large autocorrelations and pairwise cross-correlations 
(both positive and negative) in the sequence of samples generated by the
circuit dynamics. Here $\left\| \b{M} \right\|_F^2 \equiv {\rm trace}(\b{MM}^\top)
=\sum_{ij} M_{ij}^2$ is the squared Frobenius norm of $\b{M}$.

\begin{figure}[t!]
\includegraphics[width=\columnwidth]{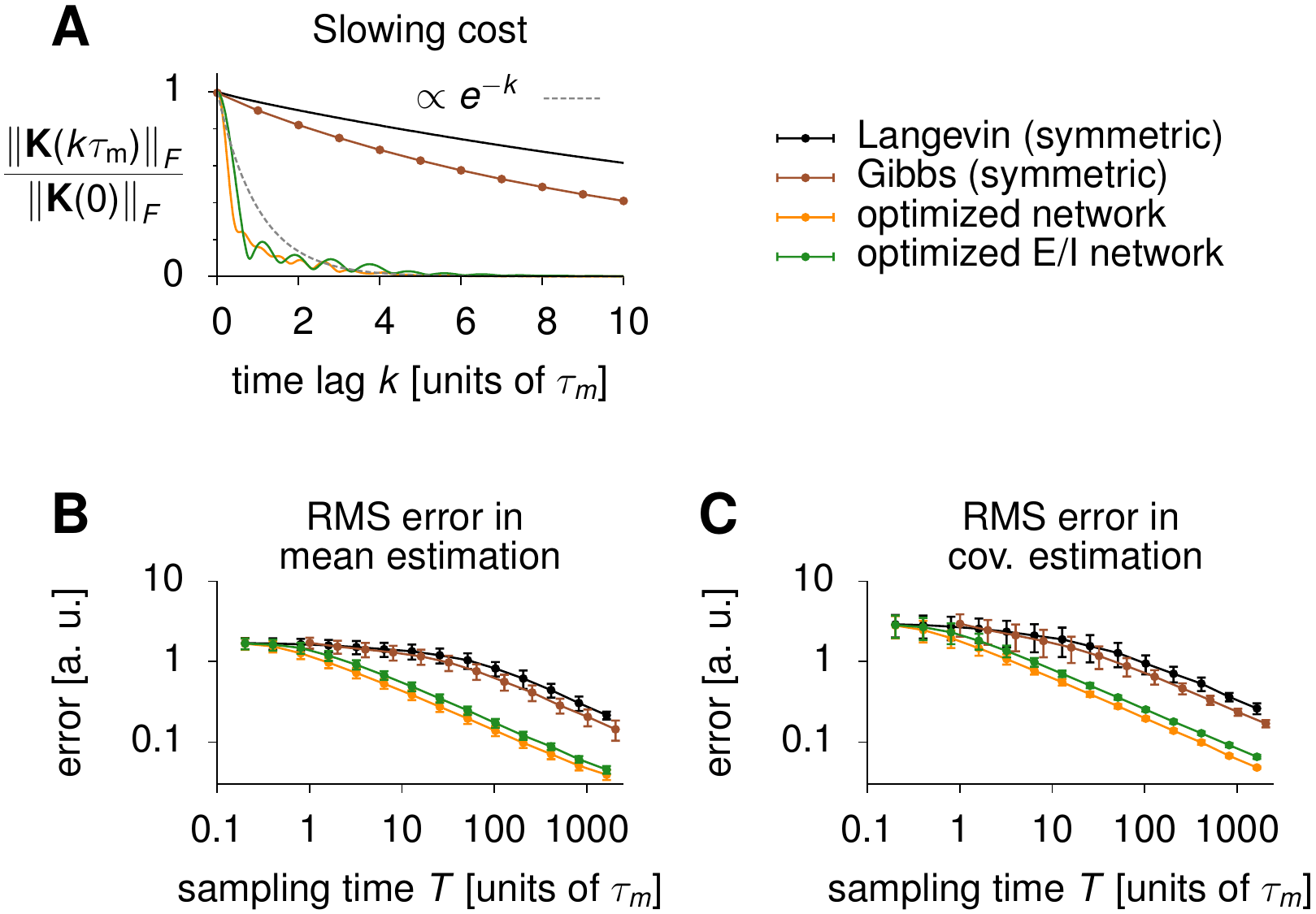}
\caption{\label{fig:cost}{\bfseries How fast is the fastest sampler?}
({\bfseries A}) Scalar measure of the statistical dependency between any two
samples collected $\tau$ seconds apart (cf. main text), for Langevin sampling
(black), Gibbs sampling (brown, assuming a full update sweep is done every
$\tau_{\rm m}$), the unconstrained optimized network (orange), and the
optimized E/I network (green). The gray line shows the behaviour of a purely
feedforward network that would merely integrate uncorrelated noise sources with
weights given by the Cholesky factor of the target $\b{\Sigma}$, and decay with
a time constant $\tau_{\rm m}$. Optimized recurrent networks do better, because
they can evolve \emph{collectively} faster than an isolated neuron would.
({\bfseries B})~Root-mean-squared error in the estimation of the posterior mean
$\boldsymbol\mu$ on the basis of a limited amount of samples $\b{r}(t)$,
collected every 2~ms during a period $T$ (x-axis, in units of $\tau_{\rm m}$).
Errorbars denote standard deviation over many trials. ({\bfseries C})~Same as
in (B), for posterior variance/covariance estimation.}
\vspace*{0.3cm}
\hrule
\end{figure}
 
\begin{figure*}[!t]
\centering
\includegraphics[width=0.75\textwidth]{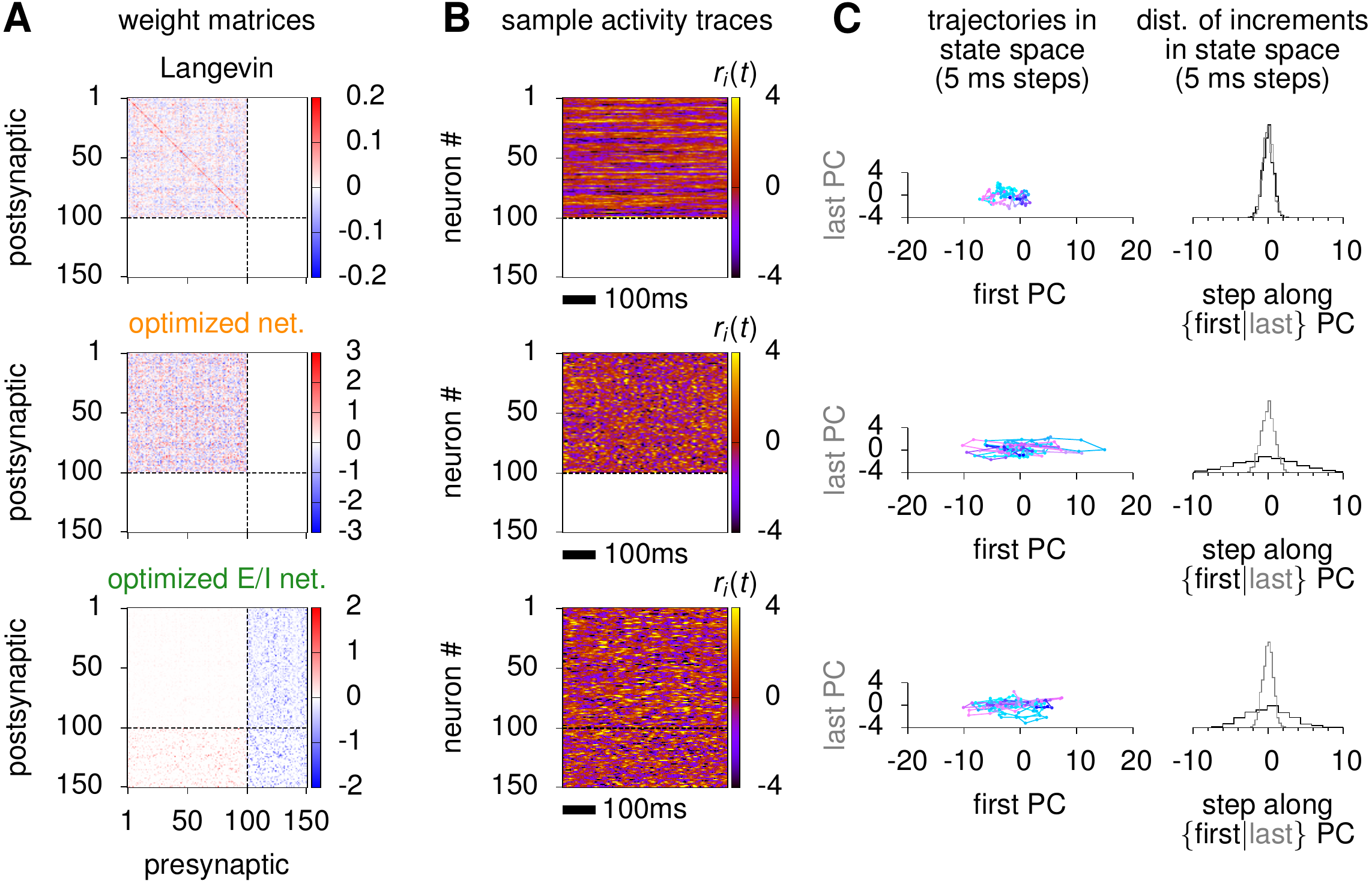}
\caption{\label{fig:samples}{\bfseries Fast sampling with optimized networks.}
{\bfseries (A)}~Synaptic weight matrices for the Langevin network (top), the 
fastest sampler (middle) and the fastest sampler that obeys Dale's law (bottom). 
Note that the synaptic weights in both optimized networks are an order of
magnitude larger than in the symmetric Langevin solution.  The first two
networks are of size $N=100$, so the last $50$ columns and rows are empty here,
but reproduced for comparison with the optimized E/I network which has size
$N=150$ (bottom).
{\bfseries (B)}~$500$~ms of spontaneous network activity ($\b{h}=0$) for
each of the three networks, for all of which the stationary distribution
of $\b{r}$ (restricted to the first 100 neurons) is the same multivariate
Gaussian. 
{\bfseries (C)}~Left: activity trajectories (the same $500$~ms as shown in (B))
in the plane defined by the topmost and bottommost eigenvectors of the
posterior covariance matrix $\b\Sigma$ (corresponding to the first and last
principal components of the activity fluctuations $\b{r}(t)$). For the E/I
network, the projection is restricted to the excitatory neurons.
Right:~distribution of increments along both axes, measured in $5$~ms time
steps. Langevin sampling takes steps of equal size along all directions,
while the optimized networks take much larger steps along the directions of
large variance prescribed by the posterior. 
}
\bigskip
\hrule
\end{figure*}

We can use this measure of slowness to illustrate the slow mixing behaviour of
Langevin sampling on a toy covariance matrix. We generate a random $\b\Sigma$
of size $N=100$ from a full, random orthonormal basis
$(\b{u}_1,\ldots,\b{u}_N)$\footnote{obtained by
Gram-Schmidt orthonormalization of a set of N random Gaussian vectors.} as:
\begin{equation}\label{eq:toysigma}
\b\Sigma = \frac{\sigma_0^2-(1-p)}p \sum_{i=1}^{pN} \b{u}_i\b{u}_i^\top
+ \sum_{i=pN+1}^N \b{u}_i \b{u}_i^\top
\end{equation}
One can easily check that the average variance, i.e. ${\rm trace}(\b\Sigma)/N$,
is equal to $\sigma_0^2$, which we set to $3$. We choose $p=0.1$, resulting in
a fairly broad distribution of pairwise correlation coefficients in $\b\Sigma$
($\sigma_r \approx 0.15$). \Cref{fig:cost}A illustrates the behaviour of
Langevin sampling by plotting $\left\| \b{K}(\b{S}=0,\tau) \right\|_{\rm F}$ as
a function of the time lag $\tau$: as predicted above in \Cref{sec:langevin},
mixing is indeed an order of magnitude slower than the single-neuron time
constant $\tau_{\rm m}$. Note that $\psi_{\rm slow}$ in \Cref{eq:slowingcost}
is proportional to the area under the squared curve in \Cref{fig:cost}A.
The slow dynamics of Langevin sampling are also illustrated in
\Cref{fig:samples}B (top), in which 500~ms of network activity are shown.

Using the same measure, we can also look at the speed of Gibbs sampling,
another widely used sampling technique (e.g.  \cite{Mezard09,Buesing11}).  It
is defined as a Markov chain that operates in discrete time, and to compare its
mixing speed with that of our linear stochastic dynamics, we assume that a
single discrete step (in which all neurons have been updated once) consumes a
time $\tau_{\rm m}$. The speed of Gibbs sampling is comparable to that of
Langevin here: samples are still very correlated on a timescale of order $\sim
10 \:\tau_{\rm m}$.

We now show that the skew-symmetric matrix $\b{S}$ can be optimized for
sampling speed, by directly minimizing the slowing cost $\psi_{\rm
slow}(\b{S})$, subject to an $L_2$-norm penalty. Our overall cost
function is therefore:
\begin{equation}\label{eq:costfunction}
\mathcal{L}(\b{S}) \quad\equiv\quad \psi_{\rm slow}(\b{S})
+ \frac{\lambda_{L_2}}{2N^2} \left\| \b{W}(\b{S}) \right\|^2_F
\end{equation}

It is well known \cite{Gardiner85} that $\b{K}(\b{S},\tau)$ obeys the following
differential equation:
\begin{equation}
\tau_{\rm m}\frac{d\b{K}(\b{S},\tau)}{d\tau} = 
  \left[\b{W}(\b{S})-\b{I}\right]\b{K}(\b{S},\tau)
\end{equation}
such that for $\tau\geq 0$
\begin{equation}
\b{K}(\b{S},\tau) = e^{\left[\b{W}(\b{S})-\b{I}\right]\:\tau/\tau_{\rm m}}\: \b{\Sigma}
\end{equation}
We may thus rewrite $\psi_{\rm slow}(\b{S})$ as
\begin{equation}
\psi_{\rm slow}(\b{S}) = \frac1{2N^2}
{\rm tr} \left [
\int_0^\infty
e^{\tau\left[\b{W}(\b{S})-\b{I}\right]} \b\Sigma^2
e^{\tau\left[\b{W}(\b{S})^\top-\b{I}\right]}
\d \tau \right]
\end{equation}
The derivatives w.r.t $\b{W}$ are given by \cite{Vanbiervliet09}:
\begin{equation}
\frac{\partial \psi_{\rm slow}(\b{S})}{\partial \b{W}} \quad=\quad
\frac{\b{QP}}{N^2}
\end{equation}
where matrices $\b{P}$ and $\b{Q}$ are the solutions of the following Lyapunov
equation pair:
\begin{align}
(\b{W}-\b{I})\b{P}+\b{P}(\b{W}-\b{I})^\top \quad&=\quad -\b\Sigma^2
\label{eq:lyapp}\\
(\b{W}-\b{I})^\top\b{Q}+\b{Q}(\b{W}-\b{I}) \quad&=\quad -\b{I}
\label{eq:lyapq}
\end{align}
These equations can be solved efficiently \cite{Bartels72}, e.g. using the
Matlab function \texttt{lyap}.  Note also that $\psi_{\rm slow}(\b{S}) = {\rm
tr}(\b{P})/2N^2$~\cite{Vanbiervliet09}.  Now, a straightforward application of
the chain rule yields
\begin{equation}\label{eq:gradientslowness}
\frac{\partial \psi_{\rm slow}(\b{S})}{\partial \b{S}} \quad=\quad
\frac1{N^2}\left[ (\b{\Sigma}^{-1}\b{PQ})^\top
- (\b{\Sigma}^{-1}\b{PQ})\right]
\end{equation}
which is skew-symmetric, as it should.  The $L_2$-penalty term in
\Cref{eq:costfunction} is more easily differentiated, yielding an overall
gradient
\begin{align}\label{eq:gradient}
\frac{\partial \mathcal{L}(\b{S})}{\partial \b{S}} \quad=\quad
&\frac1{N^2}\left[
(\b{\Sigma}^{-1}\b{PQ})^\top - (\b{\Sigma}^{-1}\b{PQ})\right]\\
 &+ \frac{\lambda_{L_2}}{N^2} \left[
\b{S\Sigma}^{-2} + \b{\Sigma}^{-2}\b{S}
\right]\nonumber
\end{align}
This gradient may be used in any gradient-based optimization approach to
minimize $\mathcal{L}(\b{S})$ and obtain the fastest regularized sampler, which
we now show on the toy covariance matrix of \Cref{eq:toysigma}.

\begin{figure}[!t]
\centering
\includegraphics[width=0.65\columnwidth]{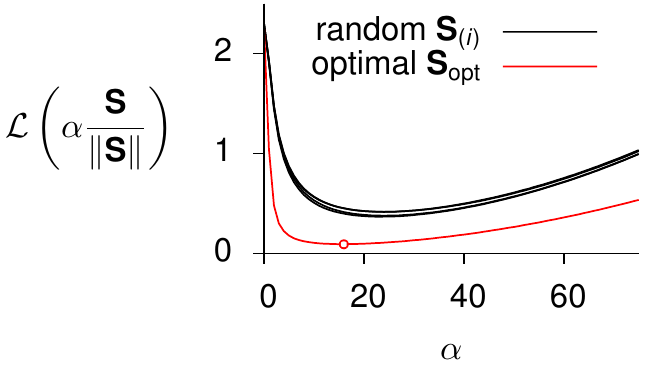}
\caption{\label{fig:convexity}The cost function $\mathcal{L}$
(\Cref{eq:costfunction}) evaluated along 4 different random directions
$\b{S}_{(i)}$ in the space of skew-symmetric matrices (black) and along the
optimal direction $\b{S}_{\rm opt}$ (red) which we found by gradient
descent. The red dot indicates the minimum of $\mathcal{L}$.}
\vspace*{0.3cm}
\hrule
\end{figure}
 
We initialized $\b{S}$ with random, weak and uncorrelated elements ($S_{i>j}
\sim \mathcal{N}(0,0.001^2)$, $S_{ji}=-S_{ij}$ and $S_{ii}=0$), and ran the
nonlinear conjugate gradient algorithm (golden section line search and
Polak-Ribi\`ere approximation) to minimize the regularized cost function in
\Cref{eq:costfunction} with $\lambda_{L_2}=0.1$. The final, optimal $\b{S}_{\rm
opt}$ induces a weight matrix $\b{W}_{\rm opt}$ given by \Cref{eq:gen:solution}
and shown in \Cref{fig:samples}A (center). Importantly, $\b{W}_{\rm opt}$ is
no longer symmetric, and its elements are an order of magnitude larger than in
the Langevin symmetric solution $\wl$. Note also that the cost function seems
to be convex along any random direction in the space of skew-symmetric matrices
(\Cref{fig:convexity}), suggesting (but not proving\footnote{For example, there
could be multiple local minima hiding on a sphere $\|\b{S} \|= {\rm constant}$.})
that $\mathcal{L}(\b{S})$ has a single minimum, and therefore that the matrix
$\b{S}_{\rm opt}$ corresponds to the fastest sampler.

The optimal sampler is an order of magnitude faster than either Langevin or
Gibbs sampling: samples are decorrelated on a timescale that is even faster
than the single-neuron time constant $\tau_{\rm m}$ (\Cref{fig:cost}A, orange).
Such decorrelation dramatically improves the sample-based estimation of the
posterior mean and covariances, as shown in \Cref{fig:cost}B and C. For
any sampler that has the right stationary distribution, the difference between
the sample estimates (samples collected every 5~ms) of $\boldsymbol\mu$ and
$\b\Sigma$ and their true values vanishes with sampling time,
\emph{asymptotically}.  For a finite number of samples, however, the estimation
error depends on how independent those samples are. For both Langevin and Gibbs
sampling, the RMS error in parameter estimation starts decaying as $1/\sqrt{T}$
(the expected asymptotic decay rate) only after $T>10$ seconds of sampling
time. In contrast, the asymptotic rate is reached by the optimal sampler after
only $\tau_{\rm m}=20$~ms, that is, from the very first sample.

We note in passing from the result of \Cref{fig:convexity} that a decent
(though sub-optimal) sampling speed can be achieved without fine-tuning,
through the use of a \emph{random} skew symmetric matrix $\b{S}$ (cf. black
curves).
  
\begin{figure}[t]
\centering
\includegraphics[width=0.65\columnwidth]{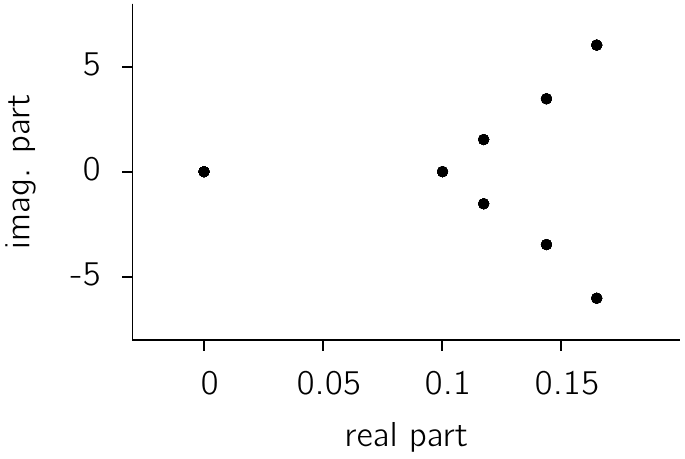}
\caption{\label{fig:eigenspectrum}{\bfseries Eigenvalue spectrum of the
connectivity matrix of the fastest sampler}. This spectrum is extremely
structured: each ``dot'' that forms the ``rotated V'' on the right is actually
made of 10 almost-identical eigenvalues, while the left-most ``dot'' is made of
30 zero eigenvalues.} 
\vspace*{0.3cm}
\hrule
\end{figure}

Perhaps intriguingly, the eigenvalue spectrum of $\b{W}_{\sf opt}$ is highly
structured~(\Cref{fig:eigenspectrum}). Moreover, the sum of its eigenvalue
squared moduli accounts for only 25\% of $\|\b{W}_{\rm opt}\|_{\rm F}^2$,
indicating $\b{W}_{\rm opt}$ is strongly nonnormal\footnote{``Nonnormal'' here
has nothing to do with ``non-Gaussian'': $\b{M}$ is nonnormal iif it is
\emph{not normal}, i.e.  $\b{M}\b{M}^\top \neq \b{M}^\top
\b{M}$.}~\cite{Trefethen05}; Indeed, for a normal matrix $\b{W}$ -- such as
the Langevin solution $\wl$ --, one would expect $\sum_i \left|\lambda_i\right|^2
= \| \b{W} \|_{\rm F}^2$.  Deviation from normality can have important
consequences for the dynamics in a linear stochastic network such as the one we
consider here: the eigenvectors of $\b{W}$ are expected \emph{not} to be
orthogonal, such that the apparent activity decay along those eigenvectors (at
a speed governed by the corresponding eigenvalue real part) can hide large
albeit transient amplification of momentary perturbations along some other
directions in state space~\cite{Murphy09,Goldman09,Hennequin12a,Hennequin14}.
It is illuminating to visualize activity trajectories in the plane defined by
the topmost and bottommost eigenvectors of $\b\Sigma$, i.e. the first and last
principal components (PCs) of the network activity (\Cref{fig:samples}C).
Inspecting these trajectories in 5~ms time steps, we see that the distribution
of discrete increments generated by Langevin sampling are identical in both
directions. Since accurate sampling requires the last PC to have small
variance (at least relative to the first PC), those distributions of increments
must be narrow, which explains the slowness of Langevin sampling: a lot of very
small steps must be taken along the first PC. In contrast, the optimal network
is not limited by the last PC, and can indeed make much larger transient
excursions along the first PC (\Cref{fig:samples}C, middle).

\section{\label{sec:ei:opt}Balanced E/I networks for fast sampling}

\newcommand\e{{\rm exc.}}
\renewcommand\i{{\rm inh.}}

We now consider more plausible network structures, namely balanced networks
made of neurons that are either excitatory or inhibitory (Dale's law). We
assume that there are $N_\e=N$ excitatory neurons, where $N$ is the dimension of
the distribution we want to sample from, and $N_\i$ inhibitory neurons whose
activity distribution is irrelevant (i.e. we regard inhibitory neurons as
auxiliary sampling variables, in the spirit of Hamiltonian Monte Carlo methods
\cite{Neal11}). In the following, we set $N_\e=100$ and $N_\i=50$.  Let
$M=N_\e + N_\i$ denote the total network size. We assume similar stochastic
dynamics as before, i.e.
\begin{equation}\label{eq:einet}
\d\b{r} = \frac{\d t}{\tau_m} \left(-\b{r}(t)+\b{W}\b{r}(t)+\b{F}\b{h}\right) +
\boldsymbol{\chi}\:\d\boldsymbol{\xi}(t) \end{equation}
where $\boldsymbol{\chi}$ is a diagonal matrix of cell type-specific input
noise variances:
\begin{equation}
\chi_{ii} = \left\{ 
\begin{array}{ll}
\chi_{\rm exc.} & {\rm if}\ i\leq N_{\rm e}\\
\chi_{\rm inh.} & {\rm otherwise.}
\end{array}
\right.
\end{equation}
Here $\chi_{\rm exc.}$ and $\chi_{\rm inh.}$ are two free parameters.

The connectivity matrix $\b{W}$ is now made of $N_\e$ positive columns followed
by $N_\i$ negative columns. This makes it difficult to apply the above approach
(\Cref{sec:optimization}) to find the fastest E/I sampler, as picking an
arbitrary skew-symmetric matrix $\b{S}$ in \Cref{eq:gen:solution}
will not yield the column sign structure of an E/I network in general.
Therefore, we no longer have a parametric form for the solution matrix manifold
on which to find the fastest network. However, with a few simple variations, we
can still formulate the problem as one of unconstrained optimization, as
explained now.

The first step is to parameterize $\b{W}$ as follows:
\begin{equation}\label{eq:wparam}
W_{ij} = \delta_{ij}\: s_j \: \exp \beta_{ij}
\end{equation}
where $s_j$ is a fixed sign that depends only on presynaptic neuron $j$
($s_j=+1$ for $j\neq N_\e$, $-1$ otherwise), and the $\beta_{ij}$'s are free
parameters. Note that we do not allow for autapses ($\delta_{ij}$ term in
\Cref{eq:wparam}).  Second, since the target posterior distribution specifies
only the $N_\e \times N_\e$ upper-left quadrant $\b\Sigma$ of the overall
covariance matrix which we denote by $\b\Lambda$, we are free to optimize over
the other quadrants. We parameterize $\b\Lambda$ by its Cholesky factor:
\begin{equation}
\b\Lambda = \b{L} \b{L}^\top, \qquad 
\b{L} \equiv \left(
\begin{array}{cc}
\b{L}_{11} & 0 \\
\b{L}_{12} & \b{L}_{22}
\end{array}
\right)
\end{equation}
where $\b{L}_{11}$ is the Cholesky factor of the posterior covariance matrix
$\b\Sigma$ (i.e. $\b\Sigma = \b{L}_{11} \b{L}_{11}^\top$), and the two matrices
$\b{L}_{12}$ and $\b{L}_{22}$ are free parameters. Note that $\b{L}_{12}$ is
a full rectangular matrix of size $N_\i \times N_\e$, while $\b{L}_{22}$ is
lower-triangular with dimensions $N_\i \times N_\i$.  Third, we incorporate the
Lyapunov equation (\Cref{eq:lyap}) as an additional constraint in our objective
function, which becomes
\begin{equation}\label{eq:cost:ei}
\mathcal{L}(\boldsymbol\theta) =
\psi_{\b\Lambda}(\boldsymbol\theta) +
\lambda_{\rm slow} \: \psi_{\rm slow}(\boldsymbol\theta) +
\lambda_{L_2} \: \| \b{W} \|_{\rm F}^2
\end{equation}
with
\begin{equation}\label{eq:costsigma}
\psi_{\b\Lambda}(\boldsymbol\theta) =
  \frac1{2M^2} \left\| 
  (\b{W}-\b{I})\b{\Lambda}
  + \b{\Lambda}(\b{W}-\b{I})^\top + \tau_{\rm m}{\boldsymbol\chi}^2 
  \right\|_F^2
\end{equation}
and $\psi_{\rm slow}$ is defined as in \Cref{eq:slowingcost} with
$\b{K}(\boldsymbol\theta,\tau) \equiv \langle \delta\b{r}_e(t)\:\delta\b{r}_e(t+\tau)^\top
\rangle$.  Here $\b{r}_e(t)$ is the vector of network activity $\b{r}$ at time
$t$, in which the inhibitory neurons' firing rates have been replaced by zeros,
i.e.  $\b{r}_e(t)=\b{Jr}(t)$ with
\begin{equation}
\b{J} = \left(\begin{array}{cc}
  \b{I} & 0\\
  0 & 0
\end{array}\right).
\end{equation}

When the cost term $\psi_{\b\Lambda}$ is zero, then the Lyapunov
equation
\begin{equation}
(\b{W}-\b{I})\b\Lambda + \b\Lambda (\b{W}-\b{I})^\top = -\tau_{\rm m}\boldsymbol\chi^2
\end{equation}
is satisfied, and therefore $\b\Lambda$ is the stationary covariance matrix of
the network activity. In particular, $\b\Sigma$ is the covariance matrix of the
excitatory neurons' activity, as wanted.

Finally, the vector $\boldsymbol\theta$ comprises all the free parameters we
have in this problem, i.e. the private noise variances $\chi_\e$ and $\chi_\i$,
all synaptic weight parameters $\beta_{ij}$, and all the relevant elements of
$\b{L}_{12}$ and $\b{L}_{22}$.

The gradients of $\psi_{\rm slow}(\boldsymbol\theta)$ can be obtained as in
\Cref{sec:optimization} (cf. also \cite{Vanbiervliet09}):
\begin{align}
\mathcal{\psi}_{\rm slow} \quad&=\quad \frac1{2N^2}\:
{\rm tr}(\b{J}\b\Sigma\b{Q}\b\Sigma)\\
\frac{\partial \psi_{\rm slow}}{\partial \b{W}} \quad&=\quad 
\frac{\b{QP}}{N^2}\\
\frac{\partial \psi_{\rm slow}}{\partial \b{L}} \quad&=\quad \frac1{N^2}
\left[ \b{J\Lambda Q} + (\b{J\Lambda Q})^\top \right] \b{L}
\end{align}
where $\b{P}$ and $\b{Q}$ solve
\begin{align}
(\b{W}-\b{I})\b{P} + \b{P}(\b{W}-\b{I})^\top &= -\b{\Lambda J \Lambda}\\
(\b{W}-\b{I})^\top \b{Q} + \b{Q}(\b{W}-\b{I}) &= -\b{J}
\end{align}

The derivatives w.r.t $\chi_\e$ and $\chi_\i$, as well as the gradients of
the other cost terms ($\psi_{\b\Lambda}$ and $\left\| \b{W} \right\|_{\rm
F}^2$), are more easily derived.

We performed nonlinear conjugate gradients to minimize the cost function
$\mathcal{L}(\boldsymbol\theta)$ in \Cref{eq:cost:ei} with
$\lambda_{L_2}=\lambda_{\rm slow} = 0.1$.  The results are presented in a
similar format as before, in the same figures (green lines).  The resulting
synaptic weight matrix is shown in \Cref{fig:samples}A (bottom), together with
a 500~ms-long activity sample (\Cref{fig:samples}B, bottom). This
Dale-compliant solution is almost as fast as the best (regularized)
unconstrained network (compare orange and green in \Cref{fig:cost}), indicating
that Dale's law -- unlike the symmetry constraint implicitly present in
Langevin sampling -- is not fundamentally detrimental to mixing speed.

\section{Discussion}

We have studied sampling for Bayesian inference in neural circuits, and
observed that a linear stochastic network is able to sample from the posterior
under a linear Gaussian latent variable model. Hidden variables are directly
encoded in the activity of single neurons, and their joint activity undergoes
moment-to-moment fluctuations and visits each portion of latent state space
with a frequency that matches the corresponding, prescribed posterior density.
To achieve this, external noise sources fed into the network are amplified by
the recurrent circuitry, but preferentially amplified along the state-space
directions that matter.

We have shown that a popular machine learning technique, namely Langevin
sampling \cite{Mackay03,Neal11,Welling11}, can be mapped onto such neuronal
network dynamics with what turns out to be an unfortunate choice of a
\emph{symmetric weight matrix}. There, an analytical argument predicts dramatic
slowing in high-dimensional latent spaces, also consistent with numerical
simulations. Samples are correlated on a timescale that extends much beyond the
single-neuron decay time constant.

When the above symmetry constraint is relaxed, a family of other solutions
opens up that can potentially lead to much faster sampling. We chose to explore
this possibility from a normative viewpoint, and optimized the network
connectivity directly for speed of sampling. The fastest sampler turned out to
be very asymmetric, nonnormal in the mathematical sense, and typically an order
of magnitude faster than Langevin sampling.

Notably, we also found that constraining each neuron to be either excitatory or
inhibitory, but not of a mixed type, does not impair the performance of the
fastest sampler but bridges the gap of biological plausibility.

Our fast samplers are capable of taking large steps in directions of large
posterior variance, and small steps in other directions. This, together with
the interpretation of Langevin sampling as a stochastic gradient ascent on the
log-posterior, suggests a link between our optimal sampling scheme and natural
gradient algorithms \cite{Amari98}, and potentially also with Riemanian Monte
Carlo sampling \cite{Girolami11}.

\section*{Acknowledgments}

This work was supported by a fellowship from the Swiss National Science
Foundation (G. H.), the Wellcome Trust (M. L. and G. H.) and the Gatsby
Charitable Foundation (L. A.).

\bibliographystyle{unsrt}
\bibliography{biblio}

\begin{thebibliography}{10}

\bibitem{Hennequin14b}
M.~Lengyel, G.~Hennequin, and L.~Aitchison.
\newblock Fast sampling in recurrent neural circuits.
\newblock In {\em Cosyne abstracts 2014}, Salt Lake City, USA, 2014.

\bibitem{Thorpe96}
S.~Thorpe, D.~Fize, and C.~Marlot.
\newblock Speed of processing in the human visual system.
\newblock {\em Nature}, 381:520–--522, 1996.

\bibitem{Knill04}
D.~Knill and A.~Pouget.
\newblock The bayesian brain: the role of uncertainty in neural coding and
  computation.
\newblock {\em Trends in Neurosciences}, 27:712--719, 2004.

\bibitem{Mackay03}
D.~{MacKay}.
\newblock {\em Information theory, inference, and learning algorithms}.
\newblock Cambridge University Press, 2003.

\bibitem{Fiser10}
J.~Fiser, P.~Berkes, G.~Orb\'an, and M.~Lengyel.
\newblock Statistically optimal perception and learning: from behavior to
  neural representations.
\newblock {\em Trends in Cognitive Sciences}, 14:119--130, 2010.

\bibitem{Berkes11}
P.~Berkes, G.~Orb\'an, M.~Lengyel, and J.~Fiser.
\newblock Spontaneous cortical activity reveals hallmarks of an optimal
  internal model of the environment.
\newblock {\em Science}, 331:83--87, 2011.

\bibitem{Trefethen05}
L.~N. Trefethen and M.~Embree.
\newblock {\em Spectra and pseudospectra: the behavior of nonnormal matrices
  and operators}.
\newblock Princeton University Press, 2005.

\bibitem{Murphy09}
B.~K. Murphy and K.~D. Miller.
\newblock Balanced amplification: A new mechanism of selective amplification of
  neural activity patterns.
\newblock {\em Neuron}, 61:635--648, 2009.

\bibitem{Gardiner85}
C.~W. Gardiner.
\newblock {\em Handbook of stochastic methods: for physics, chemistry, and the
  natural sciences}.
\newblock Berlin: Springer, 1985.

\bibitem{Neal11}
R.~Neal.
\newblock {MCMC} using hamiltonian dynamics.
\newblock {\em Handbook of Markov Chain Monte Carlo}, page 113–162, 2011.

\bibitem{Welling11}
M.~Welling and Y.~W. Teh.
\newblock {B}ayesian learning via stochastic gradient {L}angevin dynamics.
\newblock In {\em Proceedings of the International Conference on Machine
  Learning}, 2011.

\bibitem{Mezard09}
M.~Mezard and A.~Montanari.
\newblock {\em Information, physics, and computation}.
\newblock Oxford University Press, 2009.

\bibitem{Buesing11}
L.~Buesing, J.~Bill, B.~Nessler, and W.~Maass.
\newblock Neural dynamics as sampling: a model for stochastic computation in
  recurrent networks of spiking neurons.
\newblock {\em {PLoS} Computational Biology}, 7:1--22, 2011.

\bibitem{Vanbiervliet09}
J.~Vanbiervliet, B.~Vandereycken, W.~Michiels, S.~Vandewalle, and M.~Diehl.
\newblock The smoothed spectral abscissa for robust stability optimization.
\newblock {\em {SIAM} J on Optimization}, 20:156–171, 2009.

\bibitem{Bartels72}
R.~H. Bartels and G.~W. Stewart.
\newblock Solution of the matrix equation {AX+XB=C}.
\newblock {\em Communications of the ACM}, 15:820--826, 1972.

\bibitem{Goldman09}
Mark~S. Goldman.
\newblock Memory without feedback in a neural network.
\newblock {\em Neuron}, 61:621--634, 2009.

\bibitem{Hennequin12a}
G.~Hennequin, T.~P. Vogels, and W.~Gerstner.
\newblock Non-normal amplification in random balanced neuronal networks.
\newblock {\em Physical Review E}, 86:011909, 2012.

\bibitem{Hennequin14}
G.~Hennequin, T.~P. Vogels, and W.~Gerstner.
\newblock Optimal control of transient dynamics in balanced networks supports
  generation of complex movements.
\newblock {\em Neuron, accepted}, 2014.

\bibitem{Amari98}
S.~I Amari.
\newblock Natural gradient works efficiently in learning.
\newblock {\em Neural computation}, 10:251–276, 1998.

\bibitem{Girolami11}
M.~Girolami and B.~Calderhead.
\newblock Riemann manifold langevin and hamiltonian monte carlo methods.
\newblock {\em Journal of the Royal Statistical Society: Series B (Statistical
  Methodology)}, 73(2):123–214, 2011.

\end{thebibliography}

\end{document}